\documentclass[
    ,final            % use final for the camera ready runs
  ]
  {aipproc}

\layoutstyle{8x11double}

\begin{document}

\title{Supercooled Liquids under Shear: Computational Approach}

\author{R. Yamamoto}{
  address={Department of Physics, Kyoto University, Kyoto 606-8502, Japan.}
  ,altaddress={PRESTO, Japan Science and Technology Agency, 
4-1-8 Honcho Kawaguchi, Saitama, Japan.} % additional visiting address
}

\author{K. Miyazaki}{
  address={Department of Chemistry and Chemical Biology, Harvard University,
MA 02138, USA.}
}

\author{D.R. Reichman}{
  address={Department of Chemistry and Chemical Biology, Harvard University,
MA 02138, USA.}
%  ,altaddress={<author1 address>} % additional visiting address
}

\begin{abstract}
We have performed molecular dynamics simulations for a model
two-dimensional soft-core mixture in a supercooled state.
The mixture exhibits a slow structural relaxation in a quiescent state,
however, the relaxation is much enhanced in sheared states.
There observed surprisingly small anisotropy both in the coherent and
incoherent density correlation functions even under extremely 
strong shear which is $10^3$ times faster than the structural 
relaxation rate.
The present simulation results agree well with predictions 
of the recently developed mode-coupling theory in shear.

\end{abstract}

\maketitle

%%%%%%%%%%%%%%%%%%%%%%%%%%%%%%%%%%%%%%%%%%%%
%% MAINMATTER
%%%%%%%%%%%%%%%%%%%%%%%%%%%%%%%%%%%%%%%%%%%%

%\section{Introduction}

As liquids are cooled toward the glass transition, 
the dynamics is drastically slowed down,
while only small changes can be detected in the static properties.
One of the main targets in theoretical investigations on 
the glass transition is to identify the mechanism of the 
drastic slowing-down.
Beside this fundamental problem, 
a striking example occurs when one brings glassy materials away 
from equilibrium, for instance, by changing temperature rapidly or 
applying shear flow to them.
There appears a variety of unique phenomena such as aging or shear thinning.
Although these phenomena are not only conceptually but also practically 
important, physical properties of glassy materials 
in nonequilibrium conditions has not yet been understood well.
A couple of years ago, we performed extensive molecular dynamics (MD)
simulations in two dimensions (2D) and three dimensions (3D) 
on binary soft core mixtures with and without shear flow.
It was found that the dynamical properties of the mixtures under shear 
can be fairly mapped onto those at quiescent states 
at higher temperatures \cite{yamamoto1998,yamamoto2002}.
We found also the surprising isotropy in the tagged particle motions 
even under extremely high shear, which may justify the simple mapping idea.
In the present study, we calculate intermediate scattering functions
by using the method proposed by Onuki \cite{onuki1979,onuki1997} 
to investigate microscopic dynamics of glassy materials in shear flow.
Simulations have been done in 2D to compare the present 
computational results directry with the theory
developed recently in 2D \cite{miyazaki2002}.

%\section{Simulation}

Our model system is composed of two different particle spices
$1$ and $2$, which interact via the soft-core potential
\begin{equation}
v_{\alpha\beta}(r)=
\epsilon (\sigma_{\alpha\beta}/r)^{12},\hspace{5mm}
\sigma_{\alpha\beta}=(\sigma_{\alpha}+\sigma_{\beta})/2,
\end{equation}
where $r$ is the distance between two particles,
and $\alpha, \beta\in 1,2$.
We take the mass ratio $m_{2}/m_{1}=2$, the size ratio 
$\sigma_{2}/\sigma_{1}=1.4$, and the numbers of particles
$N_{1}=N_{2}=5000$.
Simulations were performed in the absence and presence of 
shear flow with being particle density and temperature
fixed at $n=(N_1+N_2)/V=0.8/\sigma_{1}^{2}$ and 
$k_BT=0.526\epsilon$, respectively.
In the sheared case, we
kept the temperature at a constant using
the Gaussian constraint thermostat
to eliminate the viscous heating effect.
Our method of applying shear is as follows:
The system was at rest for $t < 0$ for a very long
equilibration time and was then sheared for  $t > 0$.
Here we added the average velocity $\dot{\gamma} y$  to the
velocities of all the particles in the $x$ direction
at $t = 0$
and afterwards maintained  the shear flow by using
the Lee-Edwards boundary condition.
Simulation data have been taken and accumulated in steady states 
which can be realized after transient states.

Figure 1 (a) shows the geometry of shear flow in the present simulation.
As shown in Figure 1 (b), shear flow with the rate $\dot{\gamma}$ 
advect a positional vector {\bf r} as
\begin{equation}
{\bf r}(t)={\bf r}+\dot{\gamma}tr_y{\bf e}_x
\end{equation}
in the time duration $t$, where ${\bf e}_\alpha$ is a unit vector in
$\alpha\in x,y$ axis.
A similar advection can be defined in Fourier space 
for a wave vector {\bf k} as
\begin{equation}
{\bf k}(t)={\bf k}-\dot{\gamma}tk_x{\bf e}_y.
\end{equation}
The above definition enable 
us to calculate Fourier component ${\bf k}$ of 
the time correlation function
\begin{equation}
C({\bf k},t)\equiv\langle A_{-{\bf k}}(0)A_{{\bf k}(t)}(t)\rangle
\end{equation}
in shear flow.
We calculated the incoherent and coherent scattering functions
by using the definitions
\begin{equation}
Fs({\bf k},t)=\frac{1}{N}{\bigg\langle}\sum_{i=1}^{N}
e^{[-i\{{\bf k}(t)\cdot{\bf r}_i(t)-{\bf k}\cdot{\bf r}_i(0)\}]}{\bigg\rangle}
\end{equation}
and 
\begin{equation}
F({\bf k},t)=\frac{1}{N}{\bigg\langle}
\sum_{i=1}^{N}
e^{[-i{\bf k}(t)\cdot{\bf r}_i(t)]}
\sum_{j=1}^{N}
e^{[i{\bf k}\cdot{\bf r}_j(0)]}
{\bigg\rangle},
\end{equation}
respectively.
To examine anisotropy in the dynamics, 
the wave vector is taken in four different 
directions ${\bf k}_{10}$, ${\bf k}_{01}$, ${\bf k}_{11}$, 
and ${\bf k}_{-11}$, where
\begin{equation}
{\bf k}_{\alpha\beta}=\frac{k_0}{\sqrt{\alpha^2+\beta^2}}(\alpha{\bf e}_x+\beta{\bf e}_y),
\end{equation}
$k_0=2\pi/\sigma_1$, and $\alpha,\beta\in 0,1$
as shown in Figure 1 (d).

\begin{figure}[t]
\includegraphics[width=0.9\linewidth]{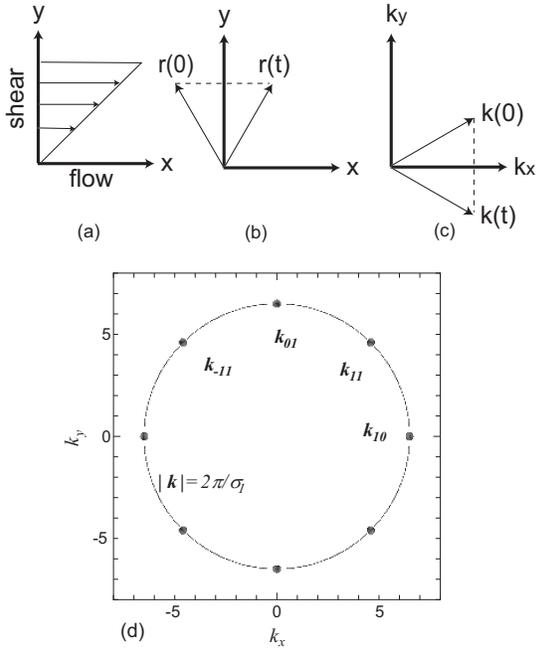}
\caption{
(a) Geometry of shear flow. (b) Shear advection in real space.
(c) Shear advection in Fourier space. (d) Sampled wave vectors (t=0).
}
\label{fig1}
\end{figure}

%\section{Results}

Figure 2 (a) and (b) show $Fs_1({\bf k},t)$ and $F_{11}({\bf k},t)$,
respectively, at $T=0.526$ with and without shear flow.
Here the subscript $1$ denotes the smaller particle component.
The so-called $\alpha$ relaxation time $\tau_\alpha$ of
the present mixture is defined by
\begin{equation}
F_{11}(k_0,\tau_\alpha)\simeq Fs_1(k_0,\tau_\alpha)=e^{-1}
\end{equation} 
in the quiescent state.
The followings have been found;
i) $Fs_1({\bf k},t)$ and $F_{11}({\bf k},t)$ behave quite similarly both
in quiescent and sheared states.
ii) Shear accelerates drastically the microscopic structural relaxation in
the supercooled state. The structural relaxation time $\tau_\alpha$
decreases strongly with increasing shear rate as $\tau_\alpha\sim
\dot{\gamma}^{-\nu}$ with $\nu\sim 1$.
iii) The acceleration in the dynamics due to shear occurs almost 
isotropically. There observed surprisingly small anisotropy in the
correlation functions even in extremely strong shear 
$\dot{\gamma}\tau_\alpha\simeq10^3$.
This simplicity in the dynamics is quite different from behaviors 
of other complex fluids such as critical fluids or polymers in shear.
Finally we note that the recent mode-coupling theory
in shear flow \cite{miyazaki2002} predicts almost the same behaviors.

\begin{figure}[t]
\includegraphics[width=0.8\linewidth]{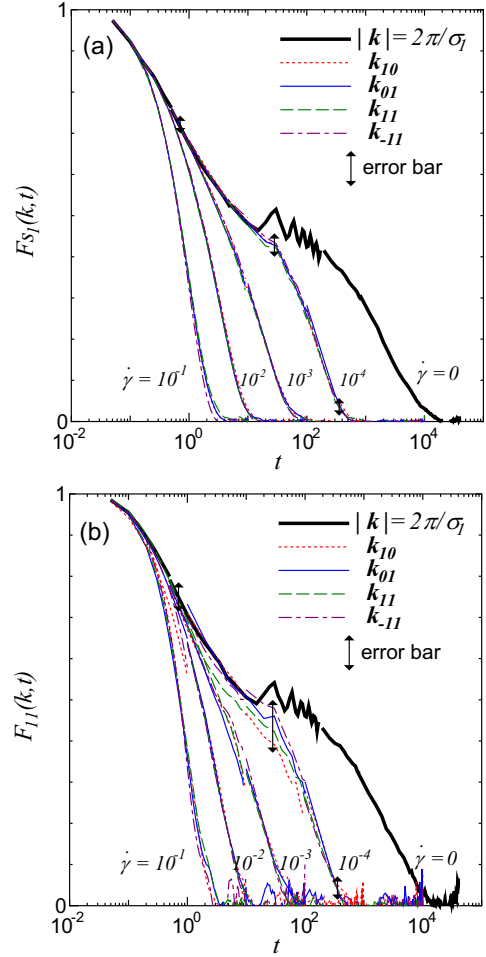}
\caption{
Intermediate scattering functions under shear; (a) incoherent part 
and (b) coherent part.
}
\label{fig2}
\end{figure}

%% If the bibliography is produced without BibTeX comment out the


\bibliography{sample}

\begin{thebibliography}{10}

\bibitem{yamamoto1998}
Yamamoto, R., and Onuki, A., {\it Phys. Rev. E} {\bf 58}, 3515-3529 (1998).

\bibitem{yamamoto2002}
Yamamoto, R., and Onuki, A., {\it J. Chem. Phys.} {\bf 117}, 2359-2367 (2002).

\bibitem{onuki1979}
Onuki, A., and Kawasaki, K., {\it Ann. Phys. (N.Y.)} {\bf 121}, 456-528 (1979).

\bibitem{onuki1997}
Onuki, A., {\it J. Phys.: Condens. Matter} {\bf 9}, 6119-6157 (1997).

\bibitem{miyazaki2002}
Miyazaki, K., and Reichman, D.R., {\it Phys. Rev. E} {\bf 66}, 050501 (2002).

\end{thebibliography}

\IfFileExists{\jobname.bbl}{}
 {\typeout{}
  \typeout{******************************************}
  \typeout{** Please run "bibtex \jobname" to optain}
  \typeout{** the bibliography and then re-run LaTeX}
  \typeout{** twice to fix the references!}
  \typeout{******************************************}
  \typeout{}
 }

\end{document}